\documentclass[twocolumn,secnumarabic, amssymb, nobibnotes, aps, prb, reprint, superscriptaddress]{revtex4-1}
\usepackage{graphicx}
\usepackage{natbib}
\begin{document}

\title{An efficient procedure for the development of optimized Projector Augmented Wave basis functions}

\author{R. J. Snow}
\affiliation{Department of Physics, University of California, Davis, California 95616, USA}
\author{A. F. Wright}
\affiliation{Sandia National Laboratories, Albuquerque, New Mexico 87185-1415, USA}
\author{C. Y. Fong}
\affiliation{Department of Physics, University of California, Davis, California 95616, USA}

\date{July 21, 2010}

\begin{abstract}

In the Projector Augmented Wave (PAW) method, a local potential, basis functions, and projector functions form an All-Electron (AE)
basis for valence wave functions in the application of Density Functional Theory (DFT).  The construction of these potentials, basis functions and projector functions
for each element can be complex, and several codes capable of utilizing the PAW method have been otherwise prevented from its use by the
lack of PAW basis sets for all atoms.  We have developed a procedure that improves the ease and efficiency of construction of PAW
basis sets.  An evolutionary algorithm is used to optimize PAW basis sets to accurately reproduce scattering properties of the 
atom and which converge well with respect to the energy cutoff in a planewave basis.  We demonstrate the procedure for the case
of Ga with the 4s, 4p, and 3d electrons treated as valence.  Calculations with this Ga PAW basis set are efficient and reproduce 
results of linearized augmented plane wave (LAPW) calculations.  We also discuss the relationship between total energy convergence with respect to the energy cutoff and the
magnitude of the matching radius of the PAW set.  

\end{abstract}
\maketitle

\section{\label{sec:Intro}Introduction}

The Projector Augmented Wave (PAW) method of Bl\"ochl\cite{PhysRevB.50.17953} provides an efficient, all-electron (AE) basis for
Density Functional Theory \cite{PhysRev.136.B864, PhysRev.140.A1133} (DFT) calculations within the frozen-core approximation.  The method
is applicable to all atoms, and the efficient treatment of first-row and transition metal elements compared with norm-conserving pseudopotential
methods is improved dramatically\cite{PhysRevB.59.1758}.  The PAW method requires the construction of a local potential, atom-centered radial (pseudo- and AE) basis functions,
and projector functions for each atom.  For PAW functions highly optimized with regard to computational efficiency, this construction can
be complex, including a local potential and multiple sets of basis and projector functions for each angular momentum channel and
principal quantum number.  To accelerate the construction of these PAW functions, and, further, to optimize with respect to the energy cutoff,
we have developed an efficient, partially automated procedure for searching this high-dimensional parameter space.

In section \ref{sec:Elements} we introduce the basic elements of the PAW and its associated parameters.  Our procedure for obtaining efficient
\emph{ab initio} PAW functions is developed in section \ref{sec:Method}.  In section \ref{sec:results} results of the construction of the
Gallium (Ga) PAW are given including logarithmic derivatives, total energy convergence,
optimization results, and lattice constant and bulk modulus calculations.  A primary key to obtaining optimal total energy convergence is a large
matching radius, and this is shown by the comparison of a series of Ga PAW basis sets with increasing matching radii.  A moderate
variation of lattice constant with the matching radius parameter is found, and this will be discussed later.  PAW basis sets constructed
in this way are expected to accurately and efficiently calculate the properties of solids within the DFT framework and the frozen-core
approximation.  

\section{\label{sec:Elements}Elements of the PAW construction}

\begin{figure}
\includegraphics[scale=0.25]{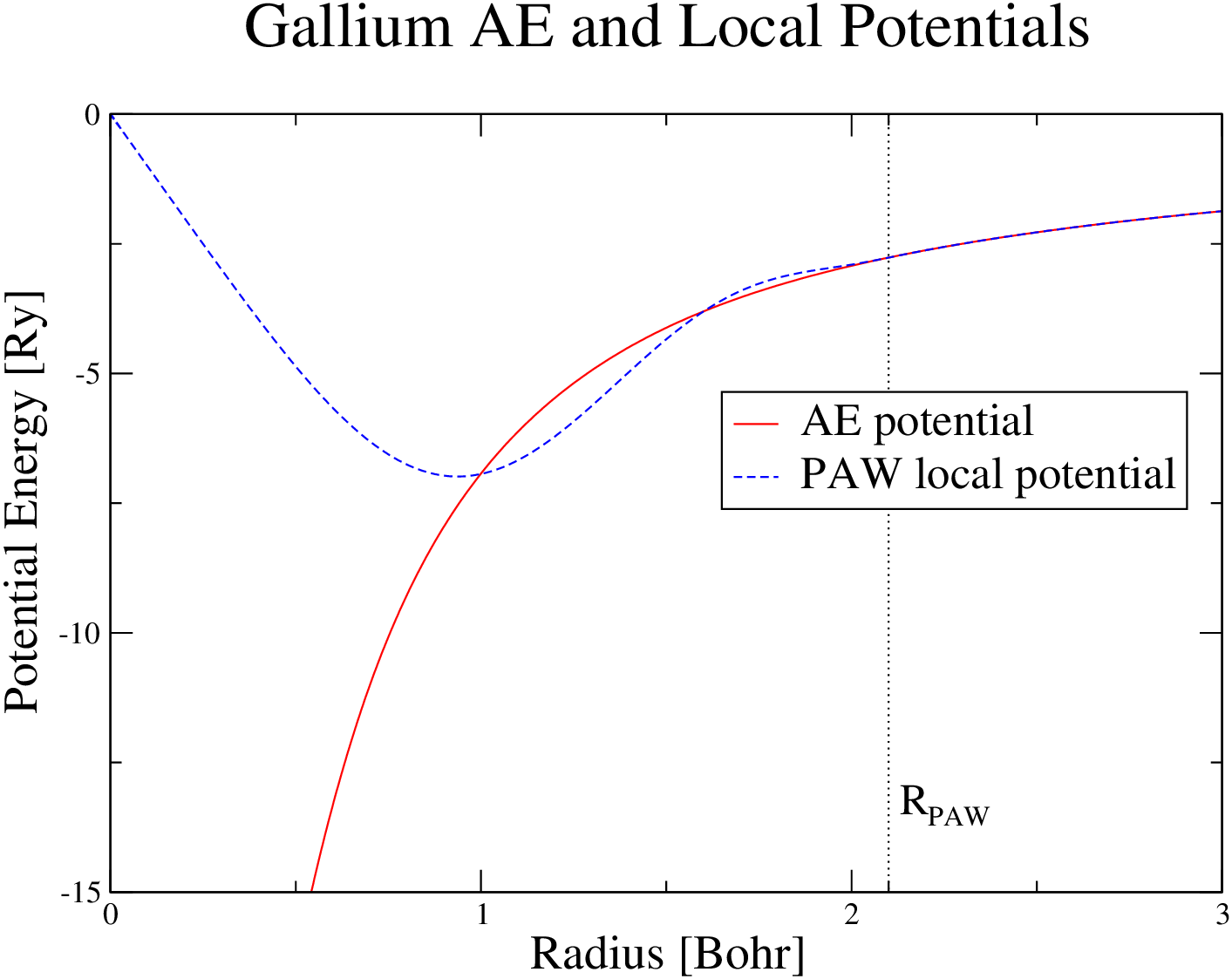}
\caption{(Color online) Ga AE and PAW local potentials.}
\label{fig:potential}
\end{figure}

\begin{figure}
\includegraphics[scale=0.3]{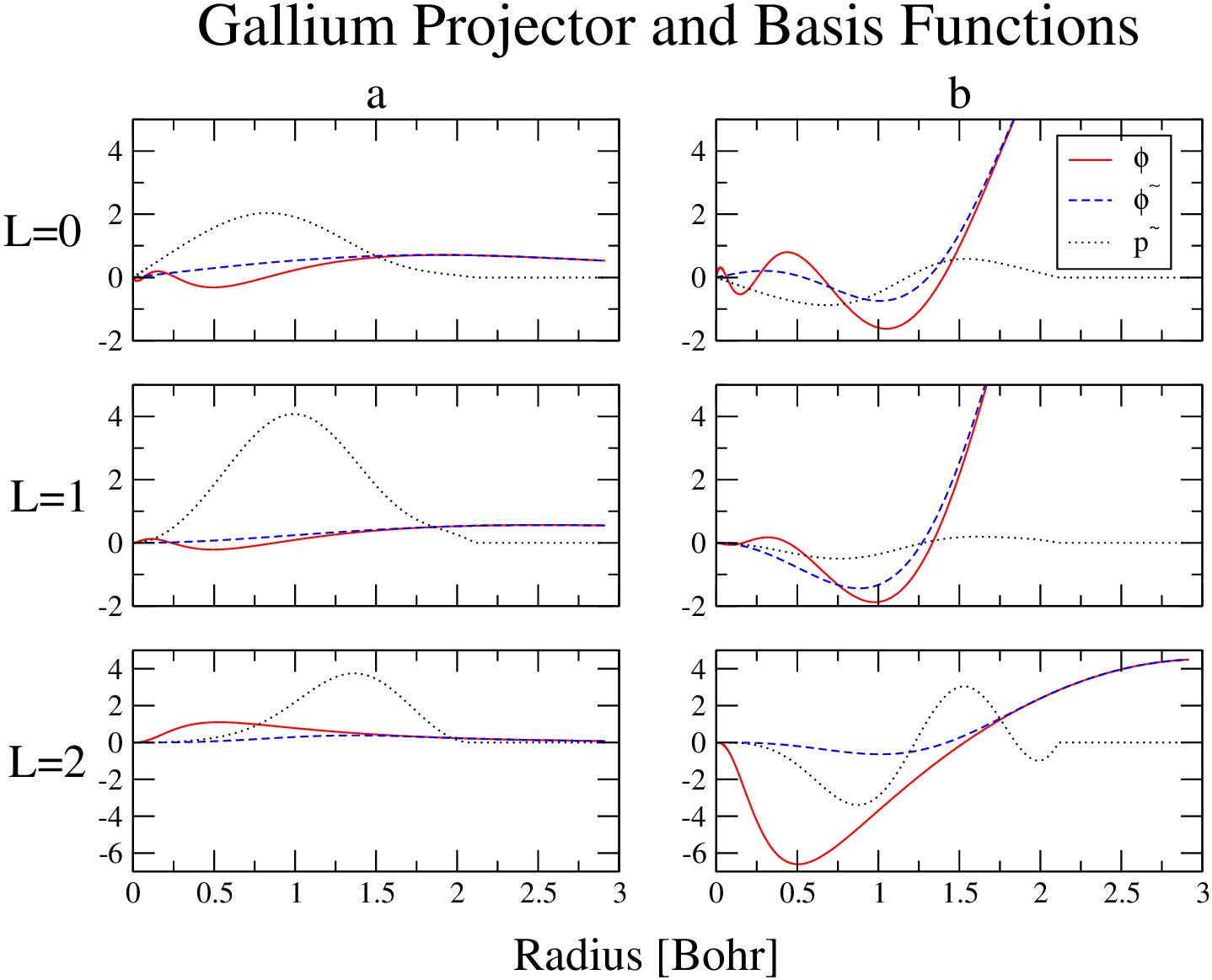}
\caption{(Color online) Ga Projector and Basis Functions.  For each angular momentum channel, two basis functions may be used.  The first PAW AE basis function,
pseudo basis function, and projector function are shown in column a and the second set of functions is in column b.}
\label{fig:projectors}
\end{figure}

The construction of PAW basis sets\cite{atompaw1} is similar to the construction of pseudopotentials.  First, a reference
configuration is chosen and electrons are divided into core and valence electrons.  A local potential is created which is smooth
within the PAW matching radius $R_{PAW}$ and matches the AE potential at and beyond $R_{PAW}$, as in figure \ref{fig:potential}.  Radial basis
functions and dual projector functions used as the basis for electronic wave functions are constructed for each angular momentum channel such
that a projection operation of $\tilde{p}$
on a smooth (pseudo) basis function $\tilde{\phi}$ reproduces the AE basis function, ${\phi}$, as in figure \ref{fig:projectors}.
A single basis function per angular momentum channel is often sufficient, but more regularly two functions per channel are required
to approach a complete basis set\cite{PhysRevB.45.11465}, and in rare cases multiple basis functions in one angular momentum channel but with
different principal quantum numbers may also be needed to represent the wave functions of semi-core electrons.  

We use the 'atompaw' program \cite{atompaw1} 
to construct basis sets in an RRKJ (Rappe, Rabe, Kaxiras, and Joannopoulos) style \cite{PhysRevB.41.1227}, one of several available
in the 'atompaw' program.  The RRKJ scheme optimizes the kinetic energy because this dominates the total energy convergence.  With RRKJ
pseudo- basis functions and projectors, each projector function is associated with an individual
matching radius, $R_{L,i}$, and an energy $E_{L,i}$.  The energy of the first pseudo basis function in each angular momentum
channel is taken as the eigenvalue of the corresponding AE solution in the atomic system, leaving three free construction parameters per angular
momentum channel when two projectors are used.  

A typical PAW will have $3*N_{L}+2$ construction parameters, including energy and matching radius parameters for the local potential, where $N_{L}$ is the
number of angular momentum channels used in the basis, usually corresponding to the angular momentum channels of the valence electrons.  These
parameters are chosen to optimize the matching of logarithmic derivatives and the total energy convergence with respect to an
energy cutoff of the planewave basis when using these PAW functions to perform calculations of solids.  Basis functions associated with
separate angular momentum channels appear to be independent with respect to logarithmic derivatives.  However, in a crystalline calculation,
these were found to have a strong interdependence in their effect on total energy convergence and in the self-consistent energy minimization.
Therefore, with a typical resolution of 0.05 Bohr in the matching radii and 0.1 Rydbergs in energy, and total ranges of 1 Bohr in the radii
and 10 Rydbergs in energies, this creates an entire parameter space of $20^{2N_{L}+1} * 100^{N_{L}+1}$ total number of possible parameter
sets, or about $1.28*10^{17}$ sets for elements with s, p, and d basis functions.  The size of this parameter space precludes a purely
brute-force attempt to optimize PAW projection basis sets.  In addition to previous advice in the literature and in various
groups\cite{MarcTorrentGuide:atompaw} regarding the selection of construction parameters, below and in section \ref{sec:Method} we discuss
further ways that this parameter space can be reduced and efficiently searched.

The matching of logarithmic derivatives of the pseudo- and AE wave functions, evaluated with respect to the radius at $R_{PAW}$ as a
function of energy, represents an equivalence of the scattering properties of the PAW basis set and the atom.  The matching of
logarithmic derivatives over a wide range of energies is one indication of a transferable basis set.  In this \emph{ab initio} construction of
PAW basis sets, the primary aim is to accurately reproduce logarithmic derivatives\cite{Heine}, and we do not optimize the sets by the
matching of results of calculations with any observable.

Traditionally PAW parameters are chosen by hand, logarithmic derivatives are inspected visually, and post-testing of basis sets for
optimal total energy convergence, and eventually for completeness by comparing with other AE results, is done individually for each constructed PAW.  This may
take many iterations to discover well-matching logarithmic derivatives which are also optimal for total energy
convergence.  Also, much of parameter space invariably remains unexplored due to its prohibitive size.  In this paper we propose an approach to reduce the time and labor associated with
producing PAW basis sets and to discover otherwise unobtainable optimized PAWs.  In our
procedure, first logarithmic derivatives are given a visual inspection to define a limited range of possible construction
parameters, thereby reducing the available parameter space significantly.  Then an evolutionary algorithm is used to further optimize both the matching of
logarithmic derivatives and the total energy convergence of a PAW.  In some cases, it is necessary to also optimize with respect to
the number of self-consistent iterations.  

\section{\label{sec:Method}Method}

The method described in this paper for obtaining an optimized PAW basis set consists of, first, the selection of configurational parameters, 
including the division of electrons into core and valence electrons and the selection of the basic types of local potentials and basis functions.
Next, parameter ranges are reduced through a visual inspection of logarithmic derivative matching and basis function smoothness.  Then, an evolutionary
algorithm further optimizes the PAW within the reduced paramter space.  Lastly, a few tests are performed to ensure the basic accuracy and tranferability of
the PAW.  Further testing of the PAW may be necessary depending on the intended calculation\cite{PhysRevB.73.035404}.

Initial parameter ranges can be restricted as follows.  A local potential matching radius $R_{PAW}$ should be as large as possible
without resulting in sphere overlap in crystalline, molecular, or molecular dynamics calculations\cite{MarcTorrentGuide:atompaw}, in 
order to optimize the total energy convergence, as will be discussed in section \ref{sec:RPAWkey}.
This is usually a little less than half the nearest neighbor distance in a particular system, with room to allow for ionic motion or
relaxation.  In our case, a Troullier-Martins style local potential is used, and the energy parameter often ranges from -2 to 5
Rydbergs, while energy parameters for the RRKJ projector functions have ranged from -2 to 8 Rydbergs.  After the local potential
matching radius $R_{PAW}$ is determined, all individual matching radii $R_{L,i}$ are
set equal to this.  Then a range for the local potential energy is determined by viewing the logarithmic derivative for the $L=L_{max}+1$
channel.  With a suitable local potential energy chosen in this range, energies of the projector functions are varied simultaneously
to find regions with good logarithmic derivatives, but with a secondary goal of obtaining smooth basis functions as well.  If well-matched
logarithmic derivatives are unobtainable with only these energy parameters, either the $R_{PAW}$ or the $R_{L,i}$ may then also be
altered to improve the logarithmic derivatives.  Ranges of parameters which yield well-matched logarithmic derivatives are then used
within the optimization program.  In the case of Ga, $E_s$ ranged from 3.0 to 7.0 Rydbergs, $E_p$ ranged from 0.0 to 7.0 Rydbergs,
$E_d$ ranged from 0.0 to 3.5 Rydbergs, and $E_{loc}$ ranged from 0.0 to 5.0 Rydbergs, leaving an effective parameter space for Ga of
about five million sets.  

In order to automate the process of optimizing logarithmic derivatives and total energy convergence, we define fitness scores for both of
these properties which will be used in an optimization algorithm.  For this purpose a numerical comparison of logarithmic derivatives
is evaluated for each potential PAW set, according to the summation, $ \sum (y_{i,AE} - y_{i,PAW} )^2 / (exp(-ABS(dy/dx)) $, where in this
expression $y_{i,AE}$ and $y_{i,PAW}$ are the all-electron and pseudo- wave function logarithmic derivatives as functions of energy, respectively.  The
purpose of the denominator here is merely to attenuate the effect of the difference between the AE and PAW values near a divergence in the 
logarithmic derivatives, as can be seen for example in figure \ref{fig:logD}.  Scores for each angular momentum
channel are summed and typically normalized by an average ideal matching score.

If logarithmic derivatives are not within a reasonable tolerance (of about 0.3), a penalty factor is given as the fitness score and returned to the
optimization routine without further testing.  Ghost states \cite{PhysRevB.44.8503, PhysRevB.41.12264} may be detected by the presence of
a divergence in the PAW logarithmic derivative where there is none
in the AE logarithmic derivative, and this case is given a large penalty factor.  If logarithmic derivatives are free of ghost states
and the matching of PAW and AE derivatives is within tolerance, crystalline tests at two successive energy cutoffs are performed, and their
energy difference, normalized by an ideal difference (about 0.01 Ry), is used as a convergence score.  Logarithmic derivative matching is emphasized with
stepped weighting, using a factor of 10 for the weight function until within an ideal tolerance (about 0.03) at which point the weight is reduced to 1,
equivalent to the weight of the total energy convergence scores.  Care must be taken so that the test with large
energy cutoff is well-converged, in order to prevent a plateau from forming in the total energy convergence figure as a function of energy
cutoff.  The Socorro\cite{socorro} DFT program with PAW functionality is used for these tests.  The normalized logarithmic derivative and
total energy convergence scores are then summed to produce a final fitness score.  

The Design Analysis Kit for Optimization and Terascale Applications \cite{Dakota} (DAKOTA) offers convenient handling of input parameters,
a parallel computation framework, and an interface to several optimization techniques and packages.  Comprehensive testing of DAKOTA
algorithms for searching PAW parameter space, or perhaps new algorithms, could reduce the time requirements for discovering optimal PAW
functions, but we find that the evolutionary algorithm of the coliny package \cite{Coliny} is sufficient.  In the coliny\_ea algorithm,
after a random group of initial PAW parameter sets are evaluated, future parameter sets are chosen as combinations of well-scoring sets
from the previous iteration.  In addition to the crossing of well-scoring parameter sets, random mutations are also applied to original
sets and to crossed sets, allowing further exploration into diverse areas in parameter
space.  In our practice, we typically use the 'two\_point' method of crossing with a 'cross\_over' rate of 0.9, a mutation rate of 0.25 with
the 'offset\_cauchy' mutation method, and a population size of 100 and maximum number of iterations anywhere from 3,000 to 25,000, depending
upon the number of parameters involved.  In the 'two\_point' method, optimal parent sets from the current generation are crossed by the 
division of parent sets into three regions, and the middle section is taken from one parent and the end sections from another parent.  In 
the 'offset\_cauchy' method for mutation, a random number is generated according to a cauchy distribution with a mean of 0, and this is
added to a selected parameter.  Since the
optimization with DAKOTA is done with nominal testing, further tests of several of the top scoring PAW basis sets are used to better gauge the total
energy convergence and to verify the transferability of the PAW.  As a function of total energy cutoff, we look at the lattice constant,
bulk modulus, and total energy convergence properties to evaluate PAW basis set optimization.

\section{\label{sec:results}Results}

\subsection{Logarithmic Derivative Matching}

\begin{figure}
\includegraphics[scale=0.3]{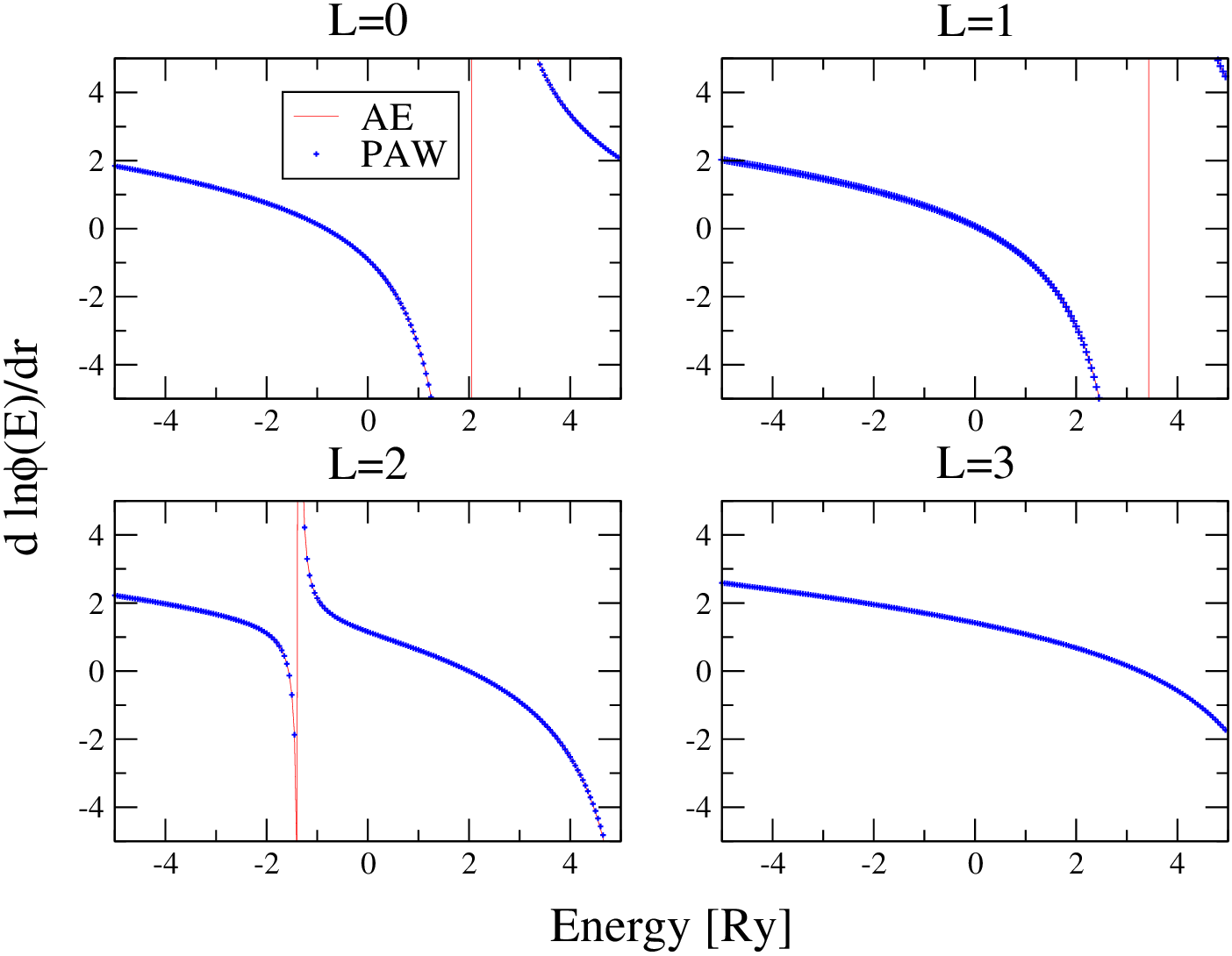}
\caption{(Color online) The AE logarithmic derivatives of the Ga atom and its PAW approximation represent the wave function matching and the scattering
properties of the atom.  Excellent agreement in all angular momentum channels L=0,1,2,3 is shown.}
\label{fig:logD}
\end{figure}

The use of an evolutionary algorithm to optimize the numerical matching of logarithmic derivatives results in
most cases with a nearly identical matching of logarithmic derivatives in the associated angular momentum channels, a feat which is
possible but painstaking when optimizing by hand.  The case of the Ga PAW is shown in figure \ref{fig:logD}.
Additionally, in many cases the matching of logarithmic derivatives can be obtained without the use of matching radii $R_{L,i}$ as
optimization parameters in each angular momentum channel; these can often be set to the local potential matching radius, $R_{PAW}$.  The energy of
expansion for a second projector function in each channel then becomes the primary parameter.  This reduces the parameter space
significantly and eliminates the problem of kinks in projector functions, saving time both in the preliminary inspection of logarithmic 
derivatives for the restriction of the parameter ranges and in the DAKOTA optimization.

\subsection{DAKOTA Total Score Optimization}

\begin{figure}[htp]
\centering
\includegraphics[scale=0.25]{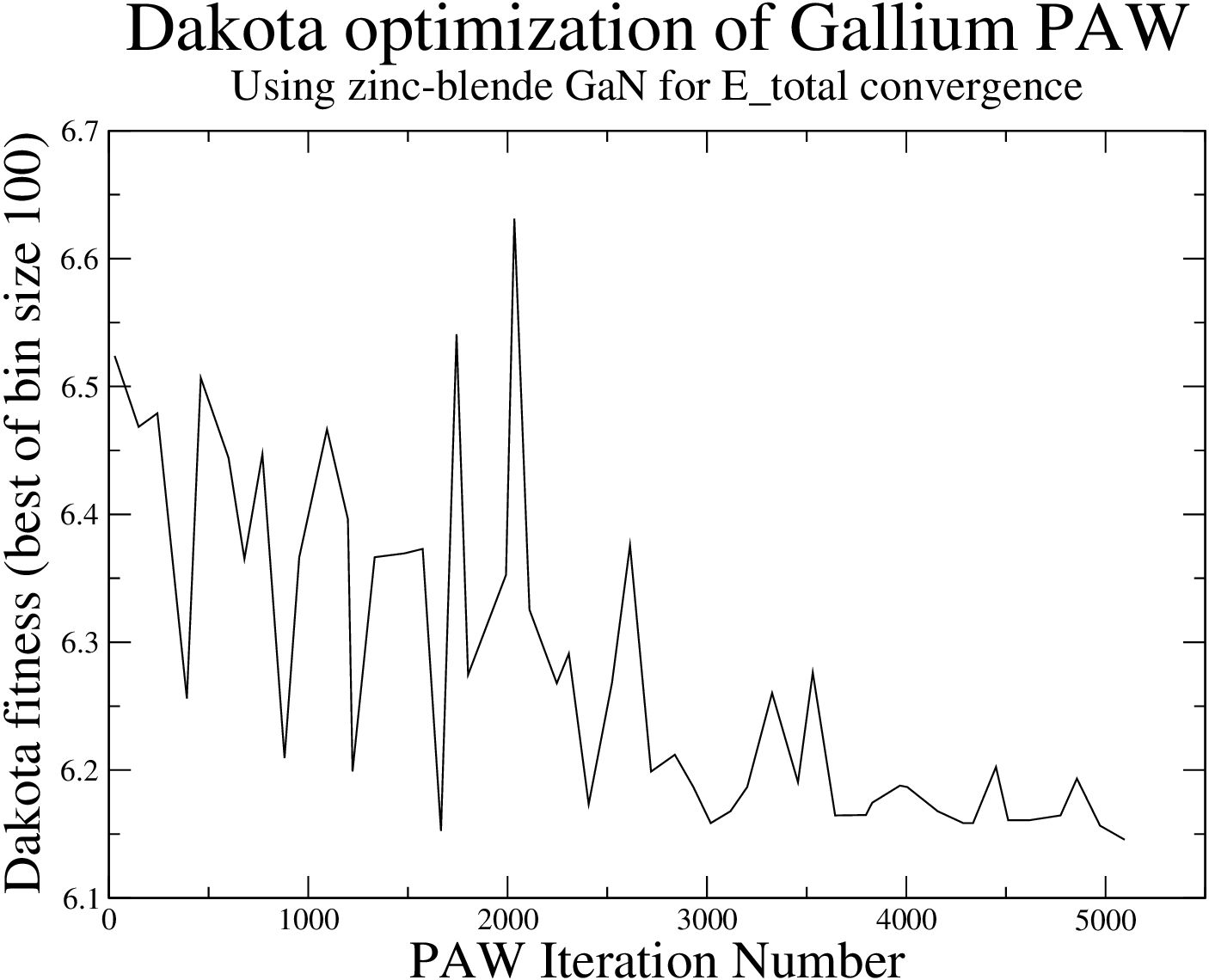}
\caption{DAKOTA total score optimization.  The best score of each group of 100 parameter sets is plotted on the y-axis.  A lower 
score represents better logarithmic derivative matching and better total energy convergence.}\label{fig:dkop}
\end{figure}

The final score used for a fitness evaluation is a combination of scores from the logarithmic
derivative matching and from the total energy convergence, as discussed above in section \ref{sec:Method}.   Figure \ref{fig:dkop} shows
the scoring of a typical DAKOTA run of around 5,000 parameter sets for the Ga PAW, using only $E_{L,i}$ with L=0,1,2 and $E_{Loc}$ as DAKOTA
parameters while keeping all $R_{L,i}$ equal to $R_{PAW}$.  If an appropriate normalization for both the logarithmic derivatives and the total
energy convergence is used, the fitness score should converge to around 1.0.  In this particular optimization, an ideal total energy convergence
of 1 mRy was used for normalizing the total energy score, the difference between the 25 Rydberg and 100 Rydberg cutoff calculations.  An
ideal logarithmic derivative matching of 0.01 was used to normalize the logarithmic derivative score.  In this case, the total energy
convergence normalization was too ambitious, and the optimal fitness score appears to converge to around 6.15, but the order of magnitude is 
acceptable.  The PAW with the best
score, known internally as Ga\_Rp2.1et001\_5095 was used in figures \ref{fig:potential}, \ref{fig:projectors}, and \ref{fig:logD}, with all
matching radii set to 2.1 Bohr, $E_s$ found to be 4.287 Rydbergs, $E_p$ found to be 4.647 Rydbergs, $E_d$ found to be 0.797 Rydbergs, and $E_{loc}$ found to
be 2.248 Rydbergs.  The standard deviation for the top ten scoring sets are 0.36, 1.79, 0.03, and 0.09 for $E_s$, $E_p$, $E_d$, and $E_loc$,
respectively.

\subsection{Variability of convergence and completeness of the Ga PAW}

\begin{figure}[htp]
\centering
\includegraphics[scale=0.25]{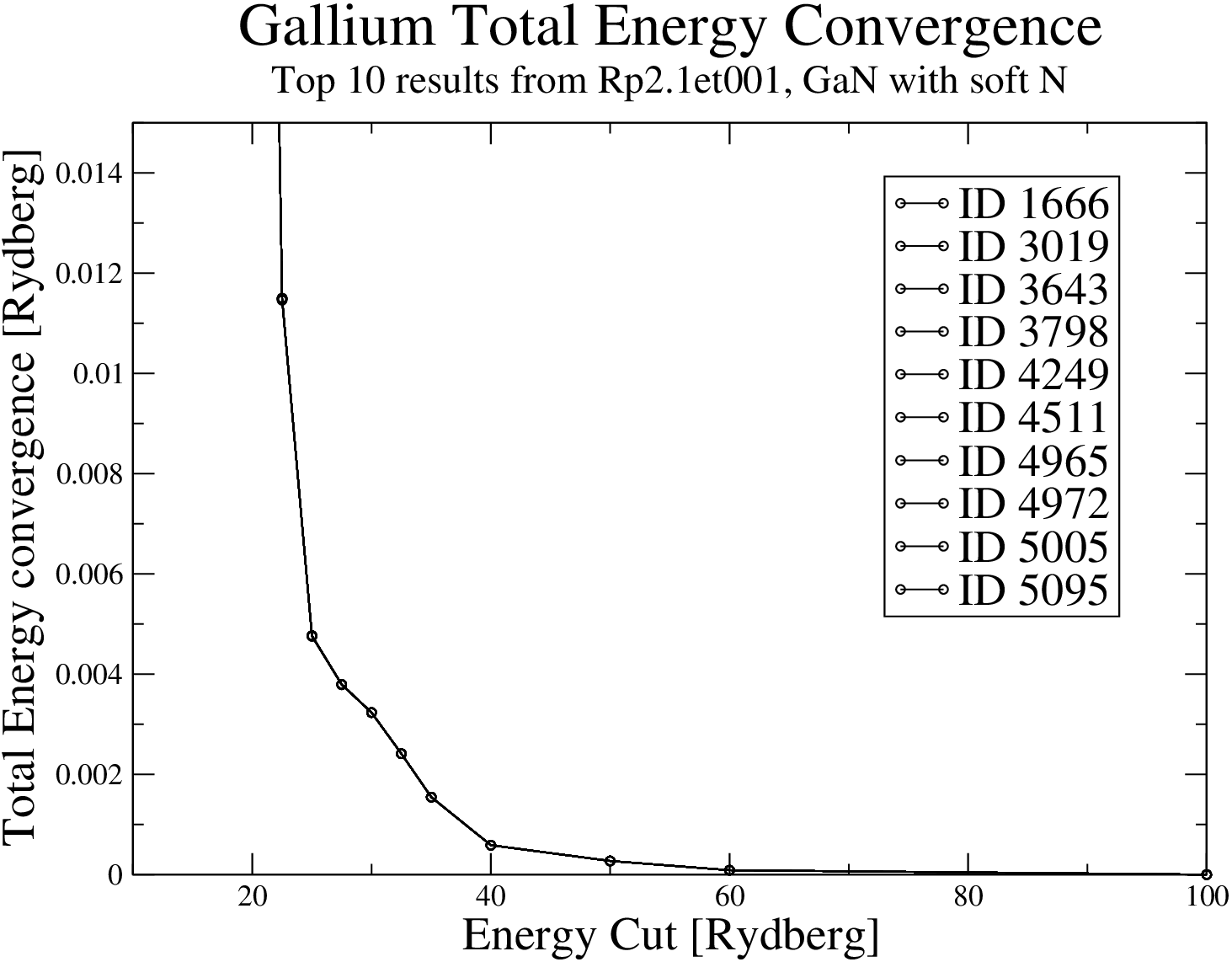}
\caption{Total energy convergence rates for the top ten results from a particular run are indistinguishable.  Matching radii $R_{L,i}$ and
$R_{PAW}$ are all equal to 2.1 Bohr.  A soft Nitrogen PAW is used here so that the total energy convergence is determined by the Ga PAW.}
\label{fig:topEtot}
\end{figure}

\begin{figure}[htp]
\centering
\includegraphics[scale=0.25]{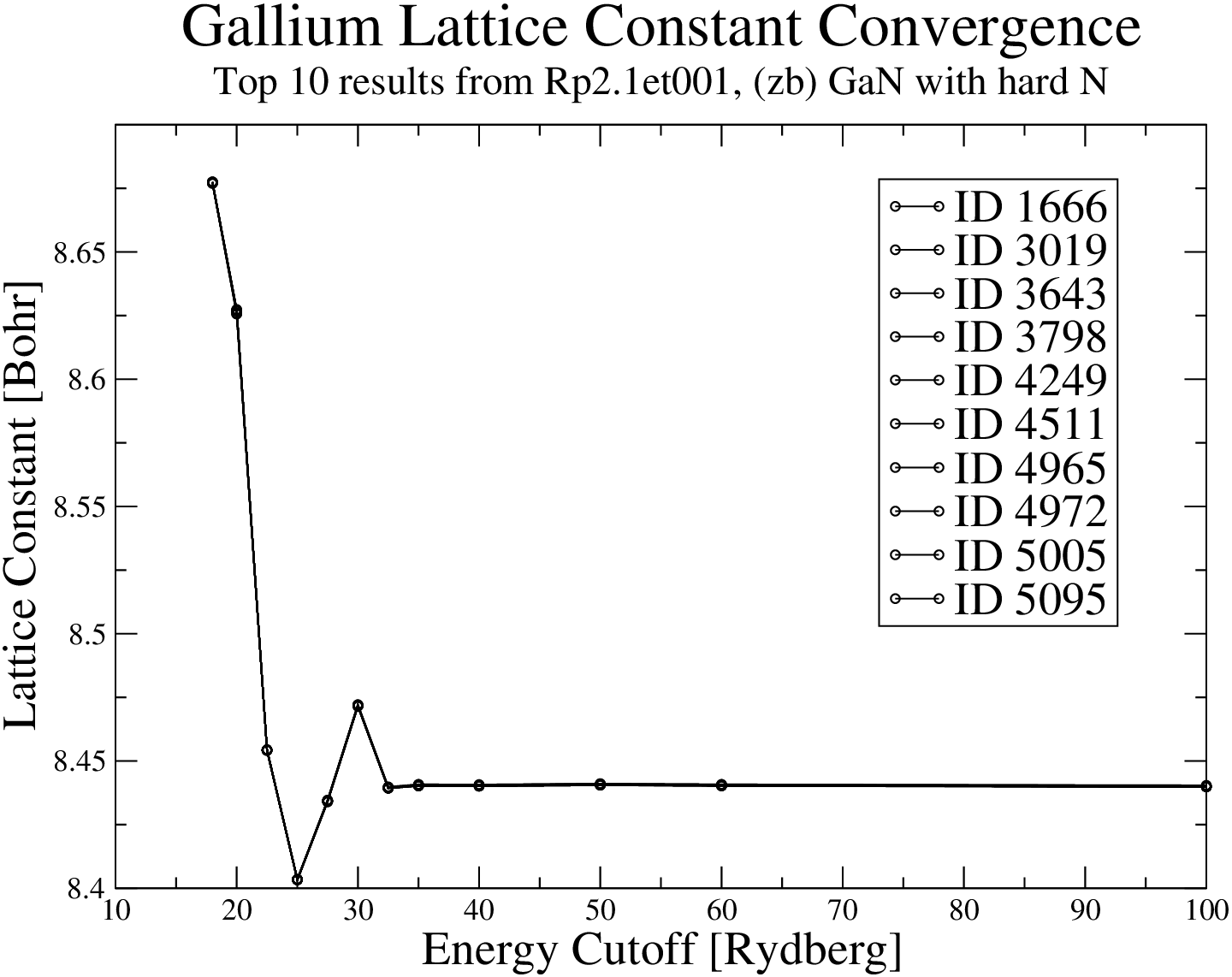}
\caption{GaN zinc-blende lattice constants as functions of energy cutoff are indistinguishable.  Matching radii $R_{L,i}$
and $R_{PAW}$ are all equal to 2.1 Bohr.  A hard Nitrogen PAW is used here so any variation in physical properties should be attributed to
Ga PAW differences.}
\label{fig:topa0}
\end{figure}

\begin{figure}[htp]
\centering
\includegraphics[scale=0.25]{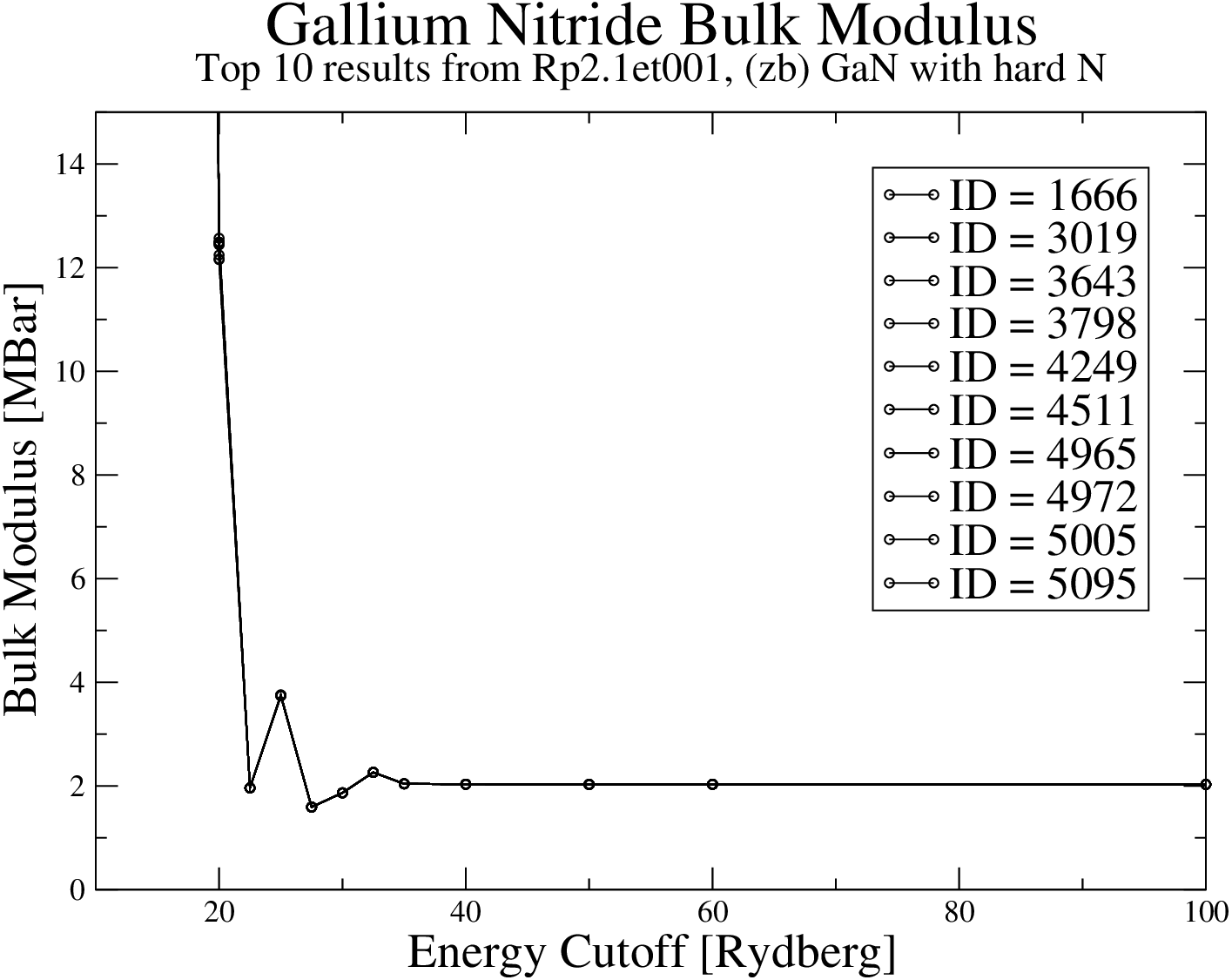}
\caption{GaN zinc-blende bulk moduli as functions of energy cutoff are virtually indistinguishable.  Matching radii $R_{L,i}$
and $R_{PAW}$ are all equal to 2.1 Bohr.  A hard Nitrogen PAW is used here so any variation in physical properties can be attributed to
Ga PAW differences.}
\label{fig:topB0}
\end{figure}

Total energy convergence rates and physical properties of PAW sets with the top ten DAKOTA scores in this trial run appear to be nearly
identical, as can be seen in figures \ref{fig:topEtot}, \ref{fig:topa0}, and \ref{fig:topB0}.  This is due partly to a large measure of
similarity in parameter values of the optimized PAW sets, but also due to the effectiveness of the PAW method and an insensitivity of
calculated properties such as lattice constants to the exact parameters of the construction.

\begin{figure}[htp]
\centering
\includegraphics[scale=0.25]{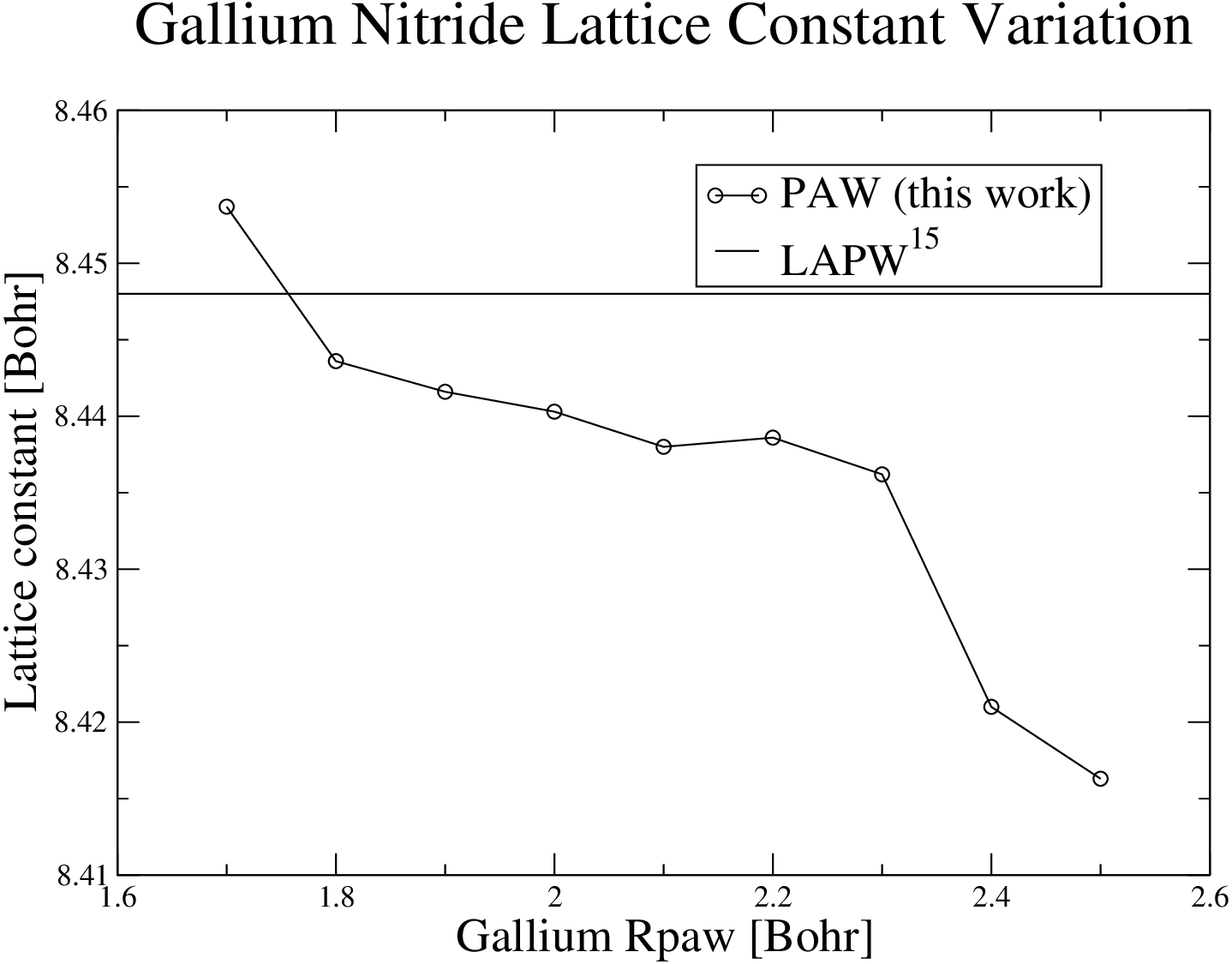}
\caption{The GaN zinc-blende lattice constant varies with the $R_{PAW}$ matching radius.  The 
N PAW used in these calculations has a matching radius of $R_{PAW}$ = 1.1 Bohr so that there is no overlap of spheres for any of the
included points.  A separate DAKOTA optimization was performed for each value of $R_{PAW}$.  The maximum difference in lattice constant
from $R_{PAW}$=1.7 to $R_{PAW}$=2.5 is less than 0.5\%.  }
\label{fig:GaLat}
\end{figure}

In the evaluation of the lattice parameter for Gallium Nitride (GaN) in the zinc-blende structure, the $R_{PAW}$ parameter for Ga had a
noticeable effect on the lattice constant, as can be seen in figure \ref{fig:GaLat}, with increasing
matching radii leading to smaller lattice constants, but with a total range of only 0.5\% in the lattice constant.  We have similarly
generated a series of N PAW sets with increasing $R_{PAW}$ parameters, but this had negligible effect on the GaN lattice constant.  

\subsection{Transferability}

Transferability in the PAW method is effective due to its AE treatment of wave functions and potentials within the core region, limited
only by the completeness of basis functions, and by the inherent limitations of the frozen-core approximation and the DFT.  To demonstrate the
transferability of the Ga PAW, we have calculated 
lattice constants and bulk moduli for zinc-blende GaN, zinc-blende GaAs, and zinc-blende GaP.  The values calculated in the local density
approximation (LDA)\cite{PhysRevB.45.13244}, shown in Table \ref{tbl:GaRp21}, are within 0.2\% error in the lattice constants in comparison
with linearized augmented plane wave (LAPW) values.  Bulk moduli are within 1\% error, except for zinc-blende GaP, which differs from
LAPW\cite{atompaw2} values by nearly 2\%.

\begin{table*}
\begin{centering}
\caption{PAW calculations of zinc-blende lattice constants and bulk moduli of GaN, GaAs, and GaP were performed in the LDA
with energy cutoffs of 100 Ry (although convergence is expected as in figures \ref{fig:topa0} and \ref{fig:topB0}).  Each of these used a
single Ga PAW, Ga\_Rp2.1et001\_5095.}
\label{tbl:GaRp21}
\begin{tabular}{|c|c|c|c|c|c|c|c|}
\hline
\hline

Crystal &PAW ID &a [Bohr] &B [MBar] &LAPW a [Bohr] &LAPW B [Bohr] &a \% error &B \% error \\
\hline
zinc-blende GaN &N.febsec5953 &8.439 &2.03 &8.447 &2.05 &-0.09 &-0.9 \\
zinc-blende GaAs &Arsenic\_d\_1058 &10.619 &0.76 &10.620 &0.77 &-0.02 &-0.7 \\
zinc-blende GaP &P.cea\_22169 &10.226 &0.91 &10.242 &0.89 &-0.16 &1.89 \\

\hline
\end{tabular}
\footnotetext{LAPW data from 'atompaw' website\cite{atompaw2}.}
\end{centering}
\end{table*}

\subsection{\label{sec:RPAWkey}$R_{PAW}$ as key to total energy convergence}

\begin{figure}[htp]
\centering
\includegraphics[scale=0.3]{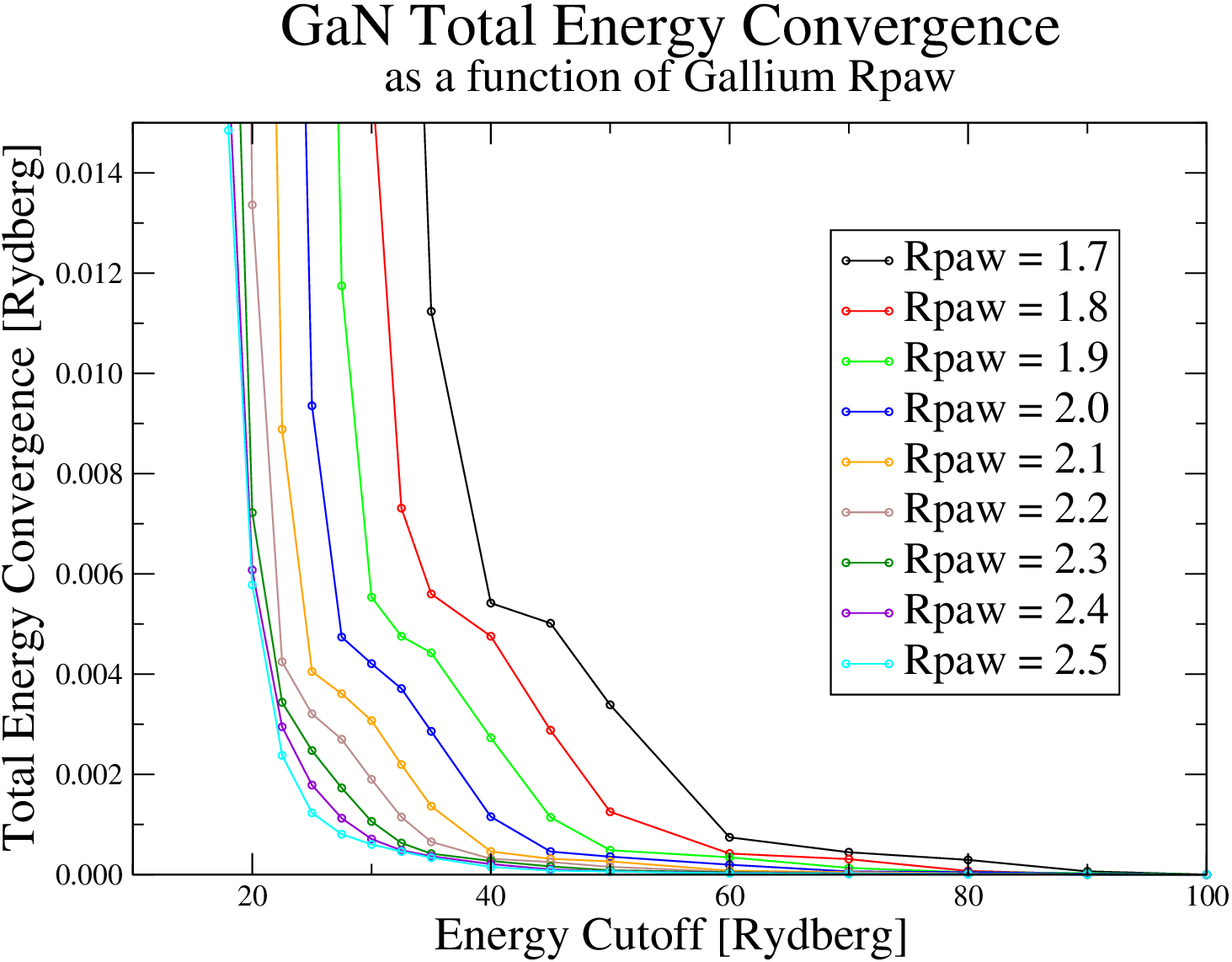}
\caption{(Color online) Correspondence between total energy convergence and the matching radius $R_{PAW}$.}
\label{fig:RpawEnergy}
\end{figure}

While the $E_{L,i}$ parameters do affect total energy convergence, the dominant bottleneck parameter is the 
local potential matching radius, $R_{PAW}$.
For a series of PAW basis sets with increasing $R_{PAW}$, figure \ref{fig:RpawEnergy} shows a
clear correspondence between the magnitude of the matching radius and the total energy convergence.  With a large matching radius, pseudo-basis
functions and the local potential can be made smooth and therefore may be expanded in a small number of plane waves.  A small amount of overlap of matching
radii of atoms in a solid calculation can have negligible effects, but in general the spheres should not overlap, limiting the size of the
matching radius and the total energy convergence.

\section{Conclusion}

The procedure described in this paper automates the process of generating optimized PAW basis sets using an evolutionary algorithm to efficiently
search a large parameter space.  In the example of the Ga PAW, the
resulting PAW matches all-electron scattering properties and is as efficient as the construction method and the particular element
permits.  Efficiency and efficacy of the PAW was confirmed in a handful of crystalline environments.  This method will be of assistance
in ongoing efforts to produce efficient PAW sets for various DFT codes.  

\section{Acknowledgements}

We acknowledge many discussions with Natalie Holzwarth and Marc Torrent, as well as for helpful guidance given on the atompaw
website\cite{MarcTorrentGuide:atompaw}.  We would like to thank Alan Tackett, Greg Walker, and Rachael Hansel for an introduction to the
DAKOTA program.  Brian Adams helped in a critical way with details of the DAKOTA program.  Sandia is a multiprogram laboratory operated by
Sandia Corporation, a Lockheed Martin Company, for the United States Department of Energy's National Nuclear Security Administration under
Contract No. DE-AC04-94AL85000.

\bibliographystyle{apsrev}

\end{document}